\documentclass[aps,
reprint,
superscriptaddress,
nofootinbib,
amsmath,amssymb,
prc,
]{revtex4-2}
\usepackage[dvipdfmx]{graphicx}
\usepackage{natbib}
\usepackage{color}
\usepackage{newtxtext}
\usepackage{dcolumn}
\usepackage{bm}
\usepackage[T1]{fontenc}

\graphicspath{{./}}
\usepackage{url}
\usepackage[colorlinks=true,allcolors=blue,breaklinks]{hyperref}
\usepackage[normalem]{ulem}

\begin{document}

\preprint{...}

\title{IMSRG-Net: A machine learning-based solver for In-Medium Similarity Renormalization Group}
\author{Sota Yoshida}
\email{syoshida@cc.utsunomiya-u.ac.jp}
\affiliation{Institute for Promotion of Higher Academic Education,~Utsunomiya University,~Mine,~Utsunomiya,~321-8505,~Japan}

\date{\today}

\begin{abstract}
We present a novel method, IMSRG-Net, which utilizes machine learning techniques as a solver for the 
in-medium Similarity Renormalization Group (IMSRG). The primary objective of IMSRG-Net is to approximate
the Magnus operators $\Omega(s)$ in the IMSRG flow equation,
thereby offering an alternative to the computationally intensive part of IMSRG calculations. 
The key idea of IMSRG-Net is its design of the loss function inspired by physics-informed neural networks
to encode the underlying {\it physics}, i.e., IMSRG flow equation, into the model.
Through training on a dataset comprising ten data points with flow parameters up to $s = 20$, capturing approximately one-eighth to one-quarter of the entire flow, IMSRG-Net exhibits remarkable accuracy in extrapolating the ground state energies and charge radii of ${}^{16}$O and ${}^{40}$Ca. Furthermore, this model demonstrates effectiveness in deriving effective interactions for a valence space.
\end{abstract}

\maketitle


\section{Introduction}
The in-medium Similarity Renormalization Group (IMSRG) method~\cite{Tsukiyama2011,Tsukiyama2012,Hergert2016Rev,StrobergRev19,Heiko_fphy20}
is a highly powerful framework to study nuclear many-body systems.
This method serves as an {\it ab initio} technique for investigating the properties of nuclei near sub-shell closures,
while also enabling the systematic derivation of effective interactions and operators for a valence space.
The IMSRG method is formulated by the unitary transformation of operators, such as the Hamiltonian, through the IMSRG flow equation.
The objective is to decouple a target subspace from the rest of many-body Hilbert space.
For ground state calculations, this entails decoupling particle-hole excitations from the reference state,
whereas for deriving effective interactions, the focus is on decoupling the valence space from the core and outside (excluded) space.
Notably, recent studies have extended the application of the IMSRG to heavier nuclei,
such as ${}^{132}$Sn and ${}^{208}$Pb~\cite{Miyagi2022Sn,Hu2022}.

Although the IMSRG method is a powerful approach, it is still computationally demanding to perform numerous calculations
for different nuclei and input nuclear interactions. Consequently, it is crucial to develop efficient methods for
conducting IMSRG calculations. The construction of such emulators or surrogate models has emerged as a prominent research topic
within the nuclear physics community, providing as a key tool for comprehending and evaluating the uncertainties 
associated with nuclear many-body calculations and realistic nuclear potentials. 
A notable example of such emulators is the eigenvector continuation (EC) method~\cite{Frame2018,Avik2021,Avik2022PRR},
which has been extensively applied to diverse nuclear many-body problems~\cite{EC_NCSM,EC_NCSM,EC_BMBPT1,EC_BMBPT2,EC_Scattering,EC_Scattering2,EC_Scattering3,EC_LEC_NCSM,EC_RMat,Franzke2022,SY_EC,Nuwan2023}.
Its significance has been recognized from a broader perspective as model order reduction~\cite{Melendez_2022,Bonilla2023}.
However, applying the EC method to IMSRG calculations poses challenges since the EC method
primarily operates on many-body wave functions (eigenvectors of a Hamiltonian),
while IMSRG calculations are performed on many-body operators, such as the Hamiltonian.

In this work, we present an alternative approach to constructing a surrogate model for IMSRG,
employing a data-driven technique based on machine learning.
The nuclear physics community has witnessed diverse applications of machine learning-based models
to replicate or assist nuclear many-body calculations (see, e.g., Refs.~\cite{RMP_MLinNP}).
We propose a machine-learning-based solver for the IMSRG flow equation, named IMSRG-Net,
which is inspired by the physics-informed neural network (PINN)~\cite{PINNRaissi,PINN_review}.
The neural network model, IMSRG-Net, is trained to approximate Magnus operators in IMSRG methods as a function of the flow parameter $s$.
The primary objective of this study is to provide a proof of concept,
exploring the potential of data-driven approaches in constructing surrogate models for IMSRG methods,
rather than aiming to accelerate the cutting-edge IMSRG calculations on supercomputers.
The test grounds for the proposed method are IMSRG calculations of ground state energies
and charge radii of ${}^{16}$O and ${}^{40}$Ca,
and derivation of effective interactions for $sd$- and $pf$-shell nuclei.

This work is organized as follows.
In Sec.~\ref{sec:METHODOLOGY}, we briefly review the IMSRG method.
Sec.~\ref{sec:IMSRGNet} is devoted to the description of the proposed method, IMSRG-Net.
The detailed procedure for training the neural network model and its computational cost are also discussed therein.
In Sec.~\ref{sec:Results}, we present the results of the IMSRG-Net calculations.
Finally, we summarize this work in Sec.~\ref{sec:Summary}.
Throughout this work, we used NuclearToolkit.jl~\cite{NuclearToolkit.jl,*Repo_NuclearToolkit.jl} for performing IMSRG calculations,
and PyTorch~\cite{PyTorch} for constructing and training the neural network model.
One can reproduce the results with the codes and data on the author's GitHub repository~\cite{Repo_IMSRGNet}.

 \begin{figure*}
      \centering{
      \includegraphics[width=16cm]{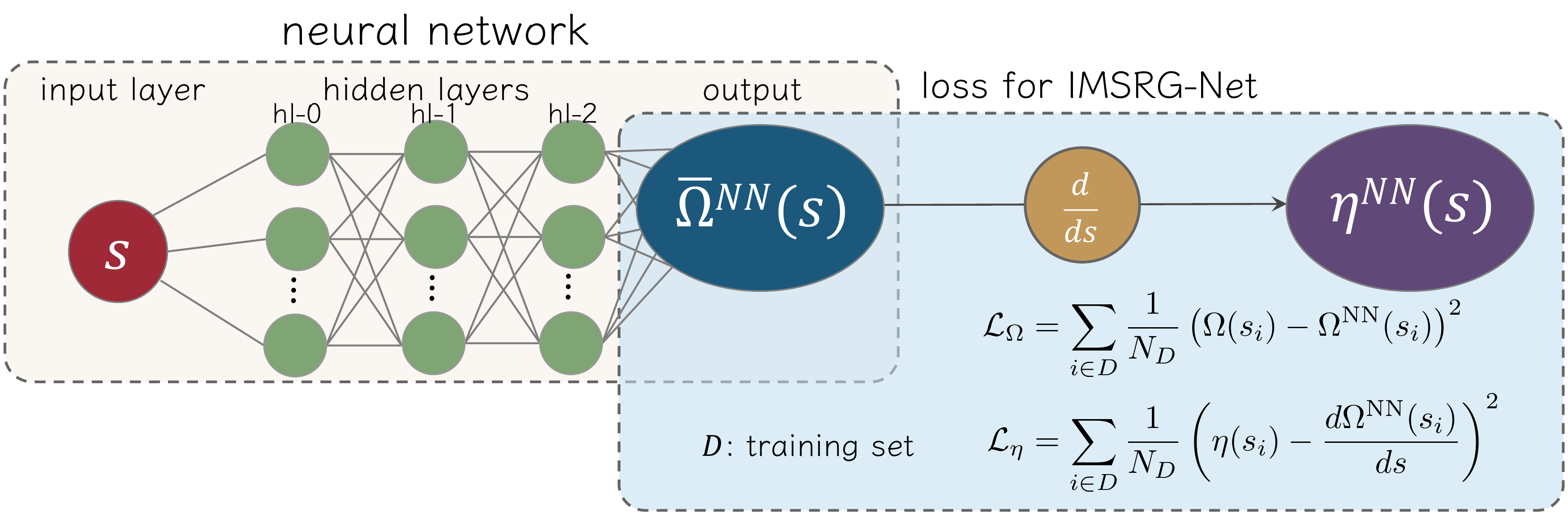}
      \caption{Schematic of IMSRG-Net. A fully connected neural network is used to generate approximated Magnus operators
      $\Omega^\mathrm{NN}(s)$ and their derivatives $\eta^\mathrm{NN}(s)$ are used as indicators of how well
      a neural network model respects the underlying IMSRG flow equation. Both the loss terms on $\Omega$
      and $\eta$ are back-propagated to train the network.
      The bar on $\Omega^\mathrm{NN}$ indicates that 
      the network output in actual calculations is {\it scaled residual} of the Magnus operator.
      See Sec.~\ref{sec:Data} for more details.
       \label{fig:IMSRGNet}}
      }
\end{figure*}


\section{METHODOLOGY \label{sec:METHODOLOGY}}
Here, we briefly review the basics of in-medium similarity renormalization group (IMSRG) method.
In IMSRG methods, one starts from a normal-ordered Hamiltonian $H(s=0)$ on a reference state,
and then performs the unitary transformation $U(s)$ to decouple particle-hole excitations from the reference state
\begin{align}
H(s) = U(s)H(0)U^{\dag}(s), 
\label{eq:Hs}
\end{align}
where $s$ is the flow parameter.
This gives the following IMSRG-flow equation:
\begin{align}
      \frac{dH(s)}{ds} & = \left[ \eta(s), H(s) \right], \label{eq:floweq} \\
      \eta(s) & \equiv \frac{dU(s)}{ds}U^\dag(s) = -\eta^\dag(s).\label{eq:eta}
\end{align}

In the last decade, 
significant progress has been made in solving the IMSRG flow equation employing the Magnus expansion~\cite{Morris_Magnus}.
Within the Magnus formulation of the IMSRG, the unitary transformation in Eq.~\eqref{eq:Hs} is explicitly evaluated by
\begin{align}
      U(s) & = e^{\Omega(s)},\label{eq:U}
\end{align}
with the anti-Hermitian Magnus operator $\Omega(s)$.
By doing this, one can write down the transformations of any operators $O(s)$ including the Hamiltonian $H(s)$ as 
\begin{align}
      O(s) = e^{\Omega(s)}O(0)e^{-\Omega(s)}. \label{eq:UOU}
\end{align}
The flow equation Eq.~\eqref{eq:floweq} is now translated into the ordinary differential equation for $\Omega(s)$ and the {\it adjoint} of $\eta(s)$
\begin{align}
      \frac{d\Omega}{ds} & = \sum^\infty_{k=0} \frac{B_k}{k!} ad^{(k)}_{\Omega}(\eta), \label{eq:adjoint} \\
      ad^{(k)}_{\Omega}(\eta) & = \left[ \Omega, ad^{(k-1)}_\Omega(\eta) \right],\\
      ad^{(0)}_{\Omega}(\eta) & = \eta,
\end{align}
where we omit the explicit dependence of $\Omega$ and $\eta$ on the flow parameter $s$, and
$B_k$ is the Bernoulli number.
One can evaluate evolved operators, Eq.~\eqref{eq:UOU}, through the Baker-Campbell-Hausdorff (BCH) formula.

Throughout this work, we employ the so-called IMSRG(2) truncation, 
where all operators are truncated up to the normal-ordered two-body (NO2B) level.
We restrict ourselves to consider the IMSRG with a single spherical reference state
and to use the so-called arctangent generator for $\eta(s)$.
The interested readers are referred to Refs.~\cite{StrobergRev19,Heiko_fphy20,Heinz2021} for additional information
on extensions of the IMSRG method, such as different basis states, generator choices, higher order contributions, and more.

In any solvers, the IMSRG flow is discretized using small finite step size $ds$,
which is usually influential on final results and thereby chosen adaptively.
Utilizing the Magnus formulation of IMSRG, one can expect the results are insensitive to the choice of the step size~\cite{Morris_Magnus}. We fixed the step size to $ds=0.25$ in this work.

As a practical approach, it is common to divide IMSRG flows into multi-steps by monitoring the norms of $\Omega(s)$
to prevent divergence and large computational time for evaluating deeply nested commutators.
In other words, the transformed Hamiltonian is used as a reference point when the norm of $\Omega(s)$
exceeds a certain tolerance. At this point, the IMSRG flow is restarted from the pivot Hamiltonian.
However, this partitioning introduces additional dependences of the final results on the chosen tolerance.
This is undesirable for the current purpose, as it obfuscates the comparison between exact IMSRG results and their approximations using the proposed method.
The issue is that the necessary number of such partitionings or splittings in exact IMSRG calculations cannot be predetermined.
To circumvent this and to regard Magnus operators as smooth functions merely on the flow parameter $s$,
we do not use such a multi-step partitioning of the Magnus operators in this work.
The only exception appears in the valence space problem (Sec.~\ref{sec:VSIMSRG}), where we adopt a two-step process:
first, the particle-hole decoupling, followed by the valence space decoupling for deriving effective interactions on a model space.
For each step, we do not use any partitioning during the IMSRG flow.

\section{IMSRG-Net\label{sec:IMSRGNet}}
Here, we describe the design of the neural network architecture and the training strategy for the proposed model, IMSRG-Net, and its computational cost.

\subsection{Design of IMSRG-Net: architecture and loss function}
As shown in Fig.~\ref{fig:IMSRGNet}, the network architecture of IMSRG-Net is simple.
The whole layers are the so-called fully connected layers (also referred to as Affine layers), which consist of the input layer, three hidden layers, and the output layer.
The network is regarded as a function to give an approximation of the Magnus operators, $\Omega^\mathrm{NN}(s)$,
for larger values of $s$ region giving converged results.
The objective of the network is to predict not observables, but the Magnus operators.
Hence, once the network is trained to predict $\Omega(s)$ accurately,
it is guaranteed that the network reproduces the IMSRG results for any observables.

As summarized in Table~\ref{tab:architecture}, the number of hidden layers is three,
and the number of nodes for each hidden layer is 48, 48, and 16 from the input side to the output side, respectively.
The adopted activation functions are the hyperbolic tangent for the first hidden layer and the softplus~\cite{Softplus} for the remaining hidden layers.

To design the loss function to be minimized, we define the following quantity:
\begin{align}
\eta^\mathrm{NN}(s) = \frac{d\Omega^\mathrm{NN}(s)}{ds}, \label{eq:eta_NN}
\end{align}
where the superscript NN represents the values obtained through the neural network model.
It should be noted that the above relation is an approximation in terms of the Magnus formulation of the IMSRG, taking the leading term of Eq.~\eqref{eq:adjoint}.
This approximation is based on the fact that the derivatives of $\Omega^{NN}$ can be evaluated easily and efficiently
by automatic differentiation or numerical differentiation.
Here we used the latter method, which is faster than the former one in our model having a large number of nodes in the output layer.

Now the loss function of IMSRG-Net is defined as the sum of the mean-squared errors (MSEs) for $\Omega^\mathrm{NN}$ and $\eta^\mathrm{NN}$:
\begin{align}
      \mathcal{L} &= \mathcal{L}_\Omega + \lambda_\eta \mathcal{L}_\eta, \label{eq:loss} \\
      \mathcal{L}_\Omega & = \sum_{i \in D} \frac{1}{N_D} \left( \Omega(s_i) - \Omega^\mathrm{NN}(s_i) \right)^2, \label{eq:lossOmega}\\
      \mathcal{L}_\eta & = \sum_{i \in D} \frac{1}{N_D} \left(  \eta(s_i) - \eta^\mathrm{NN}(s_i) \right)^2, \label{eq:lossEta}
\end{align}
where $D$ is the training data set and $N_D$ is the number of data.
The parameter $\lambda_\eta$ is introduced to balance the two terms.
The typical size of $\mathcal{L}_\eta$ can be different from that of $\mathcal{L}_\Omega$.
From our investigations, a rule of thumb to make training stable is to set $\lambda_\eta = 10^1 \sim 10^3$,
which makes the contribution of Eq.~\eqref{eq:lossEta} comparable to Eq.~\eqref{eq:lossOmega}.
We use a fixed value, $\lambda_\eta =10^2$, throughout this work.

Under this design of the loss functions, IMSRG-Net can be regarded as a special case of physics-informed neural networks (PINNs)~\cite{PINN_review,PINNRaissi}.
Each component of $\eta$ contains information regarding the channels to be decoupled through the IMSRG flow.
Such physics information is encoded through the $\mathcal{L}_\eta$ term,
serving as a soft constraint that guides the network to learn the underlying principle -- the IMSRG flow equation.

\subsection{Hyperparameters}

\begin{table}[b]
    \caption{The architecture of IMSRG-Net. The size of output layer is given by dimension of the vectorized operator, which depends on $e_\mathrm{max}$. \label{tab:architecture}}
    \begin{ruledtabular}
    \begin{tabular}{cccc}
    Layer & Activation  & Number of nodes & Bias term \\
    \hline
     hl-0 & Tanh & 48 & Yes \\
     hl-1 & Softplus & 48 & Yes  \\
     hl-2 & Softplus & 16 & Yes  \\
     output & Identity & Dim. $\Omega$ vector & No
    \end{tabular}
    \end{ruledtabular}
\end{table}

We explored various combinations of hyperparameters for neural network models.
(i).~optimizers: stochastic gradient descent (SGD), Adam~\cite{Adam}, AdamW~\cite{AdamW}, L-BFGS~\cite{LBFGS},
(ii).~activation functions: Tanh, softplus, ReLU, Leaky ReLU, ELU,
(iii).~number of hidden layers: 1, 2, 3, 4,
(iv).~number of nodes in hidden layers: 4, 8, 16, 32, 64,
(v).~parameters for optimizers (learning rate, weight decay, momentum, etc.)

Naturally, these trials do not cover all the possible combinations of hyperparameters.
Thus, it is important to remark that the architecture and hyperparameters adopted in this work
represent only one of the viable choices that can achieve the desired accuracy for the target problems.

Regarding activation functions, one of the popular choices today is the rectified linear unit (ReLU),
particularly in deep neural network models for e.g.~image recognition tasks,
due to its computational efficiency and ability to mitigate the vanishing gradient problem.
However, in our case, we found that smooth activation functions perform better than ReLU.
This may be attributed to the fact that $\Omega(s)$ is a smooth function of $s$.
For this reason, we employed the hyperbolic tangent (Tanh) function for the first hidden layer.
The Softplus function, employed in the other hidden layers, achieves both smooth outputs and avoiding the vanishing gradient problem.


We employed the AdamW optimizer~\cite{AdamW} to train the network, updating the model parameters
through the back-propagation based on the loss function Eq.~\eqref{eq:loss}.
Empirically, we observed that the model exhibits superior extrapolation performance when using AdamW
compared to the Adam optimizer with the same hyperparameters.
This may originate from the fact that the weight decay term in AdamW prevents the overfitting.
During each epoch, we did not use the entire dataset at once, but rather used a subset of the training data.
This approach, known as mini-batch learning, is a common technique to accelerate the learning process and to prevent overfitting.
Specifically, we set the batch size equal to the number of data points, which is also referred to as online learning.
By combining weight decay in AdamW and mini-batch learning (online learning),
we achieved extrapolation without sacrificing generalization ability.

The number of nodes in the hidden layers is fixed to 48 for upstream layers and 16 for the downstream layer.
While the former one is almost irrelevant to the generalization ability of our model,
we found that a larger value around 48 speeds up the training process.
On the other hand, the number of nodes in the last hidden layer closest to the output layer 
significantly influences both accuracy and computational cost.
Generally, a larger value yields better accuracy on the training, validation, and test sets,
but a smaller value is preferable in terms of computational efficiency.
We found that the sixteen is a good compromise between accuracy and VRAM usage.

The hyperparameters discussed above are kept fixed throughout the study
to demonstrate the robustness of the proposed model to different target nuclei and input potentials.

\subsection{Data and normalization \label{sec:Data}}

The operators of interests, $\Omega$ and $\eta$, are vectorized and divided into the three distinct data sets:
10 training data $\{ s= 17.75, 18.0,..., 20.0\}$, 5 validation data $\{s= 16.5, ..., 17.5\}$, and test data $\{s=\infty\}$.
Here, $s=\infty$ corresponds to the point at which IMSRG(2) converged, with $||\eta(s)|| < \epsilon_\mathrm{norm}$ as the
convergence criterion $\sim 10^{-6}$, which is one employed in IMSRG codes~\cite{imsrgcode,NuclearToolkit.jl}.
The training set is used to train the network, while the test data is used to assess the accuracy of the extrapolation by the model.
Using the validation data, we select the model that minimizes the sum of the training loss Eq.~\eqref{eq:loss} and 
MSE of the validation set across all epochs as the best model.
The number of training data points was determined by compromise between the number of data and the memory cost.
As expected, using data points closer to $s=\infty$ naturally leads to higher accuracy and vice versa.
However, in this study, the maximum value of $s$ used as training data was set to 20.
This value typically corresponds to the first 1/8 to 1/4 of the entire IMSRG flow.
In this case, the results are not significantly affected by a particular number of data points 
as long as the number of training data exceeds $\sim 5$.

In our numerical calculations, $\Omega$ values in network predictions are defined as residuals relative to a reference value $\Omega(s_t)$
such that $s_t$ represents the largest value in the training set, which is $s_t = 20.0$ in the present study.
This works as a normalization between different problems and justifies the consideration of multiple tasks using the same architecture.
Additionally, the residuals of the $\Omega$ are multiplied by a global scaling factor.
Normalizing high-dimensional vectors solely by the standard deviation of the data can sometimes lead to overflow issues.
To address this, we have determined that the scaling factor, chosen as $C_\Omega = \min\{1/\mathrm{std}(\Omega), 10^3\}$, is suitable for the current model and target nuclei.
It is important to note that the optimal scaling factor may vary depending on the specific problem,
and an inappropriate choice can result in both overflow and underflow during calculations.
To summarize, the network approximates the scaled residuals $\overline{\Omega}^\mathrm{NN}(s) \simeq C_\Omega(\Omega(s) - \Omega(s_t))$,
and those are re-scaled and added the offset through
\begin{align}
\Omega^\mathrm{NN}(s) = \overline{\Omega}^\mathrm{NN}(s) / C_\Omega + \Omega(s_t=20),
\end{align}
to be compared with exact IMSRG data.

\subsection{Computational cost for IMSRG and IMSRG-Net\label{sec:Time}}

\begin{table}[b]
      \caption{The number of elements and size of operator as a function of $e_\mathrm{max}.$ \label{tab:emaxdim}}
      \begin{ruledtabular}
      \begin{tabular}{cccc}
      $e_\mathrm{max}$ &  & $\#$ of element & size(Float32) [MB] \\
      \hline
       4 & & 34,334     & 0.1 \\
       6 & & 467,076    & 1.8 \\
       8 & & 3,526,724  & 13  \\
      10 & & 18,345,624 & 70  \\
      12 & & 73,842,358 & 282 \\
      14 & & 246,457,592& 940 \\
      \end{tabular}
      \end{ruledtabular}
\end{table}

Here, we discuss the computational time for full IMSRG(2) calculation and IMSRG-Net,
and estimate potential benefits of surrogating some part of IMSRG-flows by IMSRG-Net.

The computational cost of IMSRG-Net is primarily determined by the size of the output layer,
which depends on the size of model space specified by $e_\mathrm{max}$ truncation.
While it is desirable to use GPU accelerators for training neural network models,
the VRAM usage often exceeds the capacity of standard GPGPUs for larger $e_\mathrm{max} > 10$.
The dimension of the operator vector is primarily determined by the number of elements
in the two-body part $\{\Omega^{(2)}_{ijkl}\}$.
More precisely, one does not need to store all matrix elements of $\{\Omega^{(2)}_{ijkl}\}$ to vectorize it because of its anti-hermitian nature.
Table~\ref{tab:emaxdim} summarizes the size of the vectorized operator as a function of $e_\mathrm{max}$.
To construct fully connected neural network, these numbers are multiplied by the number of nodes in the hidden layer closest
to the output layer, which governs the size of the weight parameters.
In addition to storing those weight parameters,
the VRAM is also used to store computational graphs, gradients, and other temporary variables needed in forward and backward propagations in neural network models.
Despite the total size of operator vectors is depending on the choice of optimizers,
VRAM usage easily exceeds 20 GB for $e_\mathrm{max}$ larger than 10.
Consequently, we limit our presented results to smaller $e_\mathrm{max}$ values: 4, 6, 8, and 10.

To simplify the discussion on computational time, let us consider the $e_\mathrm{max}=10$ cases.
For full IMSRG(2) calculations for ${}^{16}$O and ${}^{40}$Ca with the EM500 interaction,
it takes about 6 hours on an i9-10940X and 4 hours on a Xeon Platinum 8360Y.
On the other hand, computation time for IMSRG-Net is independent of target nuclei and input interactions as long as the number of epochs is fixed.
It takes 2.3 hours on an NVIDIA RTX A4000 and 1.2 hours on an NVIDIA A100.
A reduction factor of computational time by IMSRG-Net, $(T_{s\leq20} + T_\mathrm{IMSRGNet})/T_\mathrm{full}$, is around 0.4-0.6 in above cases,
where $T_{s\leq20}$ denotes computational time for IMSRG(2) calculations up to $s=20$ to generate training data.
It should be noted that the reduction factor for the proposed method can vary,
because both the current IMSRG code in NuclearToolkit.jl~\cite{NuclearToolkit.jl} and the 
training strategy for neural network models may have rooms for improvement.


\begin{figure*}
\centering{
      \includegraphics[width=18cm]{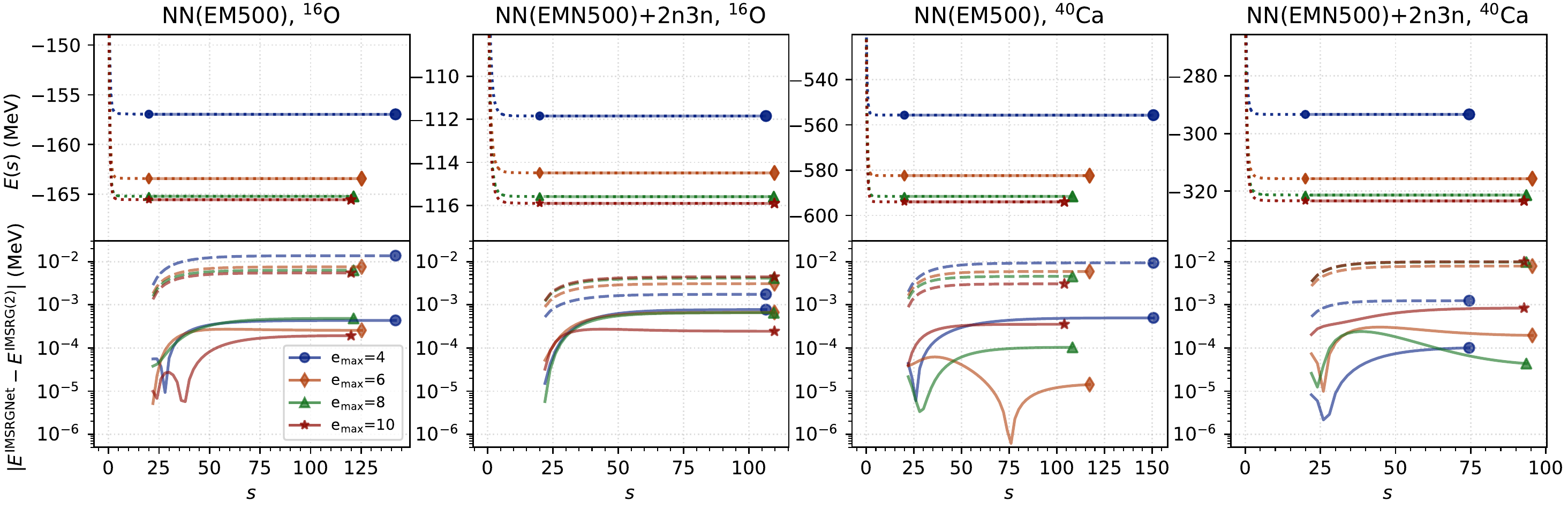}
      \caption{Ground state energies of ${}^{16}$O and ${}^{40}$Ca obtained using IMSRG(2) and IMSRG-Net.
      In the upper panels, the dashed lines correspond to the results of IMSRG(2). while the solid lines are the results of IMSRG-Net (for $s>20.0$).
      The markers are placed on the right edge of the lines to represent the points where IMSRG(2) calculations converged.
      The lower panels illustrate the energy differences between IMSRG(2) and IMSRG-Net.
      The solid lines represent the results of IMSRG-Net with $\lambda_\eta=10^2$,
      and the dashed ones represent results with $\lambda_\eta=0$.
      Further information regarding the input potentials, NN(EM500) and NN(EMN500)+2n3n, can be found in the main text.
      \label{fig:energy}}
}
\end{figure*}

\begin{table*}
      \caption{The ground state energies and charge radii of ${}^{16}$O and ${}^{40}$Ca by IMSRG(2) and IMSRG-Net.
               The $s=\infty$ corresponds to the one giving converged value of IMSRG(2). \label{tab:O16Ca40_EandRch}}
      \begin{ruledtabular}
      \begin{tabular}{llccccccccc}
            & &  &\multicolumn{4}{c}{Energy (MeV)}  & \multicolumn{4}{c}{$R_\mathrm{ch}$ (fm)} \\
       \cline{4-7}\cline{8-11}
       &  & &\multicolumn{2}{c}{$s=20$} & \multicolumn{2}{c}{$s=\infty$}  & \multicolumn{2}{c}{$s=20$} & \multicolumn{2}{c}{$s=\infty$}  \\
       \cline{4-5}\cline{6-7}\cline{8-9}\cline{10-11}
      target & interaction & $e_\mathrm{max}$  &IMSRG(2) & IMSRG-Net & IMSRG(2) & IMSRG-Net & IMSRG(2) & IMSRG-Net & IMSRG(2) & IMSRG-Net \\
      \hline
      ${}^{16}$O & EM500 &  4 & -156.9474 & -156.9474 & -156.96{\bf 11} & -156.96{\bf 07} & 2.2578 & 2.2578  & 2.261{\bf 2} & 2.261{\bf 0}  \\
            &&  6 & -163.4079 & -163.4079 & -163.415{\bf 3} & -163.415{\bf 0} & 2.2526 & 2.2526  & 2.254{\bf 7} & 2.254{\bf 6} \\
            &&  8 & -165.1876 & -165.187{\bf 5} & -165.19{\bf 32} & -165.19{\bf 27} & 2.2482 & 2.2482  & 2.249{\bf 9} & 2.249{\bf 7} \\
            && 10 & -165.5309 & -165.5309 & -165.535{\bf 9} & -165.535{\bf 7} & 2.2469 & 2.2469  & 2.248{\bf 5} &  2.248{\bf 4} \\
            & EMN500+2n3n & 4 & -111.8453 & -111.8453  & -111.84{\bf 70} & -111.84{\bf 62} & 2.3600 & 2.3600 & 2.360{\bf 7} & 2.360{\bf 5}  \\
            &&  6 & -114.4895 & -114.4895  & -114.49{\bf 25} & -114.49{\bf 18} & 2.3681 & 2.3681 & 2.369{\bf 2} & 2.369{\bf 0}  \\
            &&  8 & -115.5894 & -115.5894  & -115.59{\bf 30} & -115.59{\bf 24} & 2.3735 & 2.3735 & 2.374{\bf 8} & 2.374{\bf 6}    \\
            && 10 & -115.9040 & -115.9040  & -115.90{\bf 79} & -115.90{\bf 82} & 2.3751 & 2.3751 & 2.3765 & 2.3765 \\
      ${}^{40}$Ca & EM500 &  4 & -555.6791 & -555.6791  & -555.68{\bf 84} & -555.68{\bf 79} & 2.5947 & 2.5947 & 2.595{\bf 9} & 2.595{\bf 8}\\
            &&  6 & -582.4293 & -582.4293  & -582.435{\bf 0} & -582.435{\bf 1} & 2.5960 & 2.5960 & 2.5967 & 2.5967 \\
            &&  8 & -591.5783 & -591.578{\bf 2}& -591.582{\bf 2} & -591.582{\bf 1} & 2.5915 & 2.5915 & 2.5920 & 2.5920 \\
            && 10 & -594.0215 & -594.0215 & -594.024{\bf 2} & -594.024{\bf 6} & 2.5890  &  2.5890  & 2.589{\bf 4} & 2.589{\bf 5}  \\
            &EMN500+2n3n 
            &   4 & -293.3474 & -293.3474  & -293.348{\bf 6} & -293.348{\bf 7} & 2.8579 & 2.8579  & 2.8581 & 2.8581  \\
            &&  6 & -315.6334 & -315.6334  & -315.64{\bf 11} & -315.64{\bf 09} & 2.9082 & 2.9082  & 2.908{\bf 9} & 2.908{\bf 8} \\
            &&  8 & -321.323{\bf 3} & -321.323{\bf 2} & -321.33{\bf 20} & -321.33{\bf 19} & 2.920{\bf 1} & 2.920{\bf 0} & 2.920{\bf 9} & 2.920{\bf 8}  \\
            && 10 & -323.35{\bf 19} & -321.35{\bf 20} & -323.36{\bf 05} & -323.36{\bf 13} & 2.9252 & 2.9252 & 2.9260 & 2.9260 \\
      \end{tabular}
      \end{ruledtabular}
  \end{table*}

\section{Results of IMSRG-Net\label{sec:Results}}

\subsection{Ground state properties}

\begin{figure*}[t]
      \centering{
            \includegraphics[width=17.2cm]{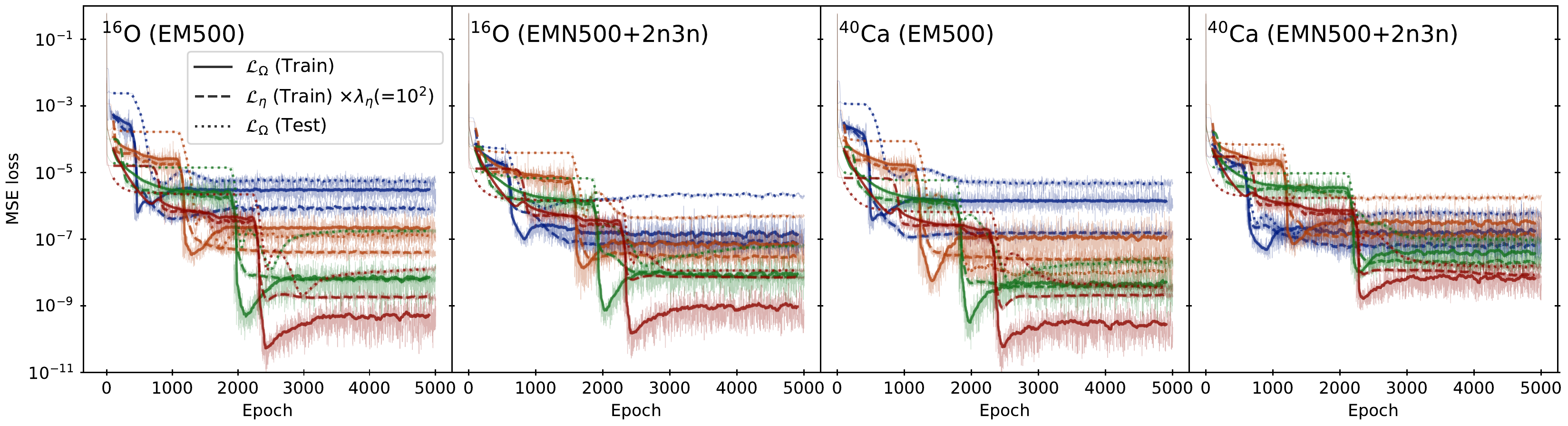}
            \caption{Learning curve of IMSRG-Net calculations.
            The lines represent moving average over 100 epochs to see the trend.
            The colors for different $e_\mathrm{max}$ truncation and abbreviations of input potentials are similar to Fig.~\ref{fig:energy}.
            \label{fig:losscurve}}
            }
\end{figure*}

In Fig.~\ref{fig:energy}, we show the results of IMSRG-Net for ${}^{16}$O and ${}^{40}$Ca using two different input interactions.
The first interaction, denoted as EM500, consists of the so-called EM nucleon-nucleon (NN) interaction~\cite{EMPRC}
regularized by a cutoff 500 MeV and softened by the similarity renormalization group (SRG) with $\lambda=2.0$ fm$^{-1}$.
The other one, denoted as EMN500+2n3n, comprises the SRG evolved EMN interaction~\cite{EKMN,EMN} up to N4LO.
The "2n3n" is representing that a density-dependent three-nucleon force~\cite{Kohno,*KohnoErratum,SY_2n3n} is added
to the softened EMN500 interaction.
Each different symbol corresponds to results with different model spaces, specified by the $e_\mathrm{max}$.

The upper panels in Fig.~\ref{fig:energy} display the results obtained by IMSRG(2) (dashed lines) and IMSRG-Net (solid lines for $s>20.0$) 
for different $e_\mathrm{max}$ truncations.
Since it is difficult to discern the differences between the two models in this scale,
we show the energy differences between IMSRG(2) and IMSRG-Net in the lower panels.
The solid lines in the lower panels are the results of IMSRG-Net, and the dashed ones represents cases where $\lambda_\eta=0$.
It is evident that incorporating the loss term on $\eta$ is crucial for training the network and achieving better extrapolation in the larger $s$ region.
The extrapolations exhibit an accuracy of less than 1 keV.

Table~\ref{tab:O16Ca40_EandRch} summarizes the ground state energies and charge radii of ${}^{16}$O and ${}^{40}$Ca evaluated by IMSRG(2) and IMSRG-Net.
The digits that differ between IMSRG(2) and IMSRG-Net are highlighted in bold.
Since we extrapolated not the energies but the Magnus operators $\Omega$ to achieve unitary transformations,
Eqs.~(\ref{eq:U}-\ref{eq:UOU}), we can compute the charge radii in a straightforward manner.
From the Table~\ref{tab:O16Ca40_EandRch}, it is apparent that the results of IMSRG-Net are in good agreement with those of IMSRG(2).
In both cases, energies and charge radii, one can see that the deviation between IMSRG(2) and IMSRG-Net
are much smaller than the residuals which are to be obtained through the rest IMSRG flow from $s=20$ to $s=\infty$.
This level of accuracy is considered sufficient to replace the remaining flow for $s > 20$ with the proposed model.

In Fig.~\ref{fig:losscurve}, the learning curves of IMSRG-Net are plotted for ground state calculations of ${}^{16}$O and ${}^{40}$Ca.
Because of using online learning, MSE losses are inevitably shaggy, more like a band rather than a line.
The corresponding moving averages of 100 epochs are plotted by the lines to see the trends of the learning curves.

As a whole, the learning curves show similar trends for different nuclei and input potentials.
It occasionally encounters a plateau, and, after a while, the MSE suddenly drops.
This trend can be understood as the effect of the AdamW optimizer.
The weight decay term gradually reduces redundant degrees of freedom in network parameters that are ineffective at reducing the loss.
Then, it leads to escaping the plateau and a subsequent drop in the training loss.
While test errors on Magnus operators are generally one to two orders of magnitude worse than those for the training sets,
the results of the observables in Tab.~\ref{tab:O16Ca40_EandRch} indicate that the extrapolation is sufficiently accurate.
Looking at the dashed lines in Fig.~\ref{fig:losscurve}, we  can see that the $\mathcal{L}_\eta$ term is also well suppressed after training
and the approximation introduced in Eq.~\eqref{eq:eta_NN} works well to inform the network of information of $\eta$.

It should be noted that the numbers (energies and charge radii) presented in this work are based on a single run specifying a random seed.
The results can vary in different environments and with different random seeds,
they can either be better or worse than the values reported here.
The uncertainty resulting from such optimization process is small,
on the same order of magnitude as the values in the boldface in Table~\ref{tab:O16Ca40_EandRch}.
To reproduce the results, one can refer to the provided repository on GitHub~\cite{Repo_IMSRGNet},
which contains the necessary information and code.

\subsection{Valence space IMSRG\label{sec:VSIMSRG}}

\begin{figure}
\centering{
      \includegraphics[width=8.6cm]{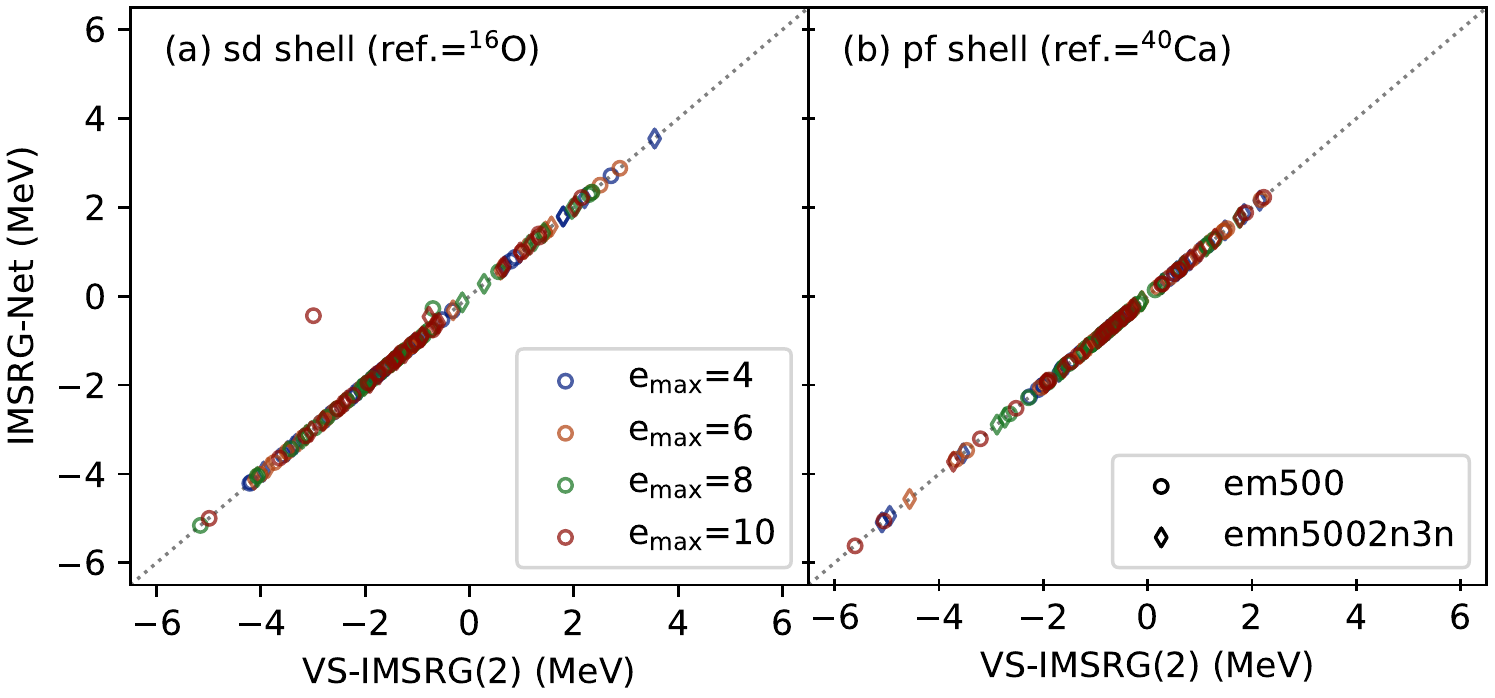}
      \caption{Correlation plot of effective interactions derived by VS-IMSRG(2) and IMSRG-Net, (a) sd shell and (b) pf shell.            
      This includes both single particle energies and two-body matrix elements, and some are randomly omitted since the number of parameters are too large to plot them all.
      The circle and diamond symbols represent the results of EM500 and EMN5002n3n, respectively.
      \label{fig:VSIMSRG}}
      }
\end{figure}

As a natural expectation, one may consider applications of IMSRG-Net to valence-space IMSRG (VS-IMSRG)
for deriving valence space effective interactions and operators.
To this end, let us consider the VS-IMSRG utilizing IMSRG-Net with the same architecture above.
The dataset is constructed simultaneously as the IMSRG case, utilizing $\Omega(s)$ up to $s=20$ as the training and validation set.
Note that the $s=\infty$ point for IMSRG(2) now serves as the starting point for the VS-IMSRG flow, i.e. $s=0$.

Fig.~\ref{fig:VSIMSRG} illustrates the correlation plot between the effective interactions derived by VS-IMSRG(2) and IMSRG-Net.
These are effective interactions on the $sd$- and $pf$-shell model spaces employing the same input interactions as mentioned above.
The reference states are chosen as the nearest doubly magic nuclei ${}^{16}$O and ${}^{40}$Ca.
As a whole, the deviation is small, which is typically less than $10^{-8}$ in mean squared error,
with a few exceptions for $e_\mathrm{max} >=8$ corresponding the symbols located far from the diagonal lines in Fig.~\ref{fig:VSIMSRG}.
These exceptions correspond to the two-body matrix elements relevant to higher orbits in the valence space,
0d3/2 for the sd shell and 0f5/2 for the pf shell.
It is found that the calculation of the commutator in the BCH formula exhibits non-convergent behavior during the VS-IMSRG flow for these cases.
For this reason, the failure in some components is attributed to not IMSRG-Net, but the numerical instability in VS-IMSRG.
When one employs the multi-step partitioning or pivoting mentioned in Sec.~\ref{sec:METHODOLOGY},
such a numerical instability can be mitigated.
However, as mentioned earlier, the question remains as to how many times partitioning can occur up to a certain flow parameter, i.e., at what point can the Magnus operator be regarded as a continuous function of $s$ and the IMSRG-Net be used to replace subsequent flows?


\section{Summary and Outlook\label{sec:Summary}}
This study explores the possibility of employing a machine-learning-based model as an alternative solver for 
in-medium similarity renormalization group (IMSRG). 
By utilizing the proposed neural network, IMSRG-Net, inspired by Physics-Informed Neural Networks (PINNs),
we achieved extrapolation with a satisfactory level of accuracy.
Energy extrapolation was accomplished with an accuracy of less than 1 keV,
and the charge radii were also well reproduced with errors of approximately 0.0001 fm.
Once accurate approximations of the Magnus operators $\Omega(s)$ are obtained,
any desired operators of interest can be evolved accurately using the Magnus formulation of IMSRG.

While we have empirically investigated the effectiveness of IMSRG-Net on various datasets,
it is important to acknowledge that there may always be exceptions and cases where this network architecture or learning strategy encounters some limitations.
One example is the issue of numerical instability within IMSRG itself, as demonstrated by the valence space results.
To further enhance the accuracy of extrapolated results and improve robustness against different target interactions and nuclei, additional inductive biases may need to be introduced.

This work may herald new approaches for IMSRG methods and offer numerous possibilities for applying PINNs-like techniques to other nuclear many-body machinery.
On the other hand, from a practical perspective on computational time,
the speeding-up effect of IMSRG calculations through IMSRG-Net is limited, as discussed in Sec.~\ref{sec:Time}.
It would be crucial to develop more computationally efficient versions of IMSRG-Net.
Since the current IMSRG-Net directly employs $\Omega(s)$ with large dimensions as the output layer of the neural network,
it becomes challenging to store and manipulate all of them on GPGPU for cases with $e_\mathrm{max}$ larger than 10.
This limitation prevents the application of IMSRG-Net for cutting-edge IMSRG calculations having larger $e_\mathrm{max}$.
For instance, working within a latent space using encoder and decoder (e.g. Ref.~\cite{DeepLFRG2022}) or constructing low-rank operator expressions while retaining maximal information through techniques like tensor decomposition could be effective.

\section*{Acknowledgments.} 
This work was supported by JSPS KAKENHI (Grants No. 22K14030) and JGC-Saneyoshi Scholarship Foundation.
This research was partly conducted using the FUJITSU Supercomputer PRIMEHPC FX1000
and FUJITSU Server PRIMERGY GX2570 (Wisteria/BDEC-01) at the Information Technology Center, The University of Tokyo.

\bibliography{references}

\end{document}